\documentclass[%
 reprint,
 amsmath,amssymb,
 aps, longbibliography
]{revtex4-1}

\usepackage{graphicx}
\usepackage{dcolumn}
\usepackage{bm}

\begin{document}

\preprint{APS/123-QED}

\title{Penetration of action potentials during collision in the median and lateral giant axons of invertebrates}

\author{Alfredo Gonzalez-Perez}
\author{Rima Budvytyte}
\author{Lars D. Mosgaard}
\author{Marius T. Stauning}
\author{S\o ren Nissen}
\author{Thomas Heimburg}
\email[]{theimbu@nbi.ku.dk}
\homepage[]{http://www.membranes.nbi.dk}
\affiliation{%
The Niels Bohr Institute, University of Copenhagen, Blegdamsvej 17, 2100 Copenhagen, Denmark\\
}%


\date{\today}

\begin{abstract}
The collisions of two simultaneously generated impulses in the giant axons of both earthworms and lobster propagating in orthodromic and antidromic direction were investigated. The experiments have been performed on the extracted ventral cords of \textit{Lumbricus terrestris} and the abdominal ventral cord of lobster, \textit{Homarus americanus}, by using external stimulation and recording. The collision of two nerve impulses of orthodromic and antidromic propagation didn't result in the annihilation of the two signals contrary to the common notion that is based on the existence of a refractory period in the well-known Hodgkin-Huxley theory. However, the results are in agreement with the electromechanical soliton theory for nerve pulse propagation as suggested by Heimburg and Jackson \cite{Heimburg2005c}. 
\end{abstract}

\pacs{Valid PACS appear here}
\maketitle


\section{introduction}
The action potential in nerves consists of a transmembrane voltage pulse of approximately 100\,mV that propagates along the neuronal axon. In 1952, Hodgkin and Huxley proposed that this pulse results from a selective voltage-dependent breakdown in membrane resistance for potassium and sodium \cite{Hodgkin1952b}. Ions flow along the concentration gradients through channel proteins modeled as electrical resistors and the Hodgkin-Huxley (HH) model is thus intrinsically dissipative.  Hodgkin compared the action potential to 'a burning fuse of gunpowder' \cite{Hodgkin1964}. Time-scales in the model, intended to describe relaxation processes in the proteins, are contained in the parametrization of the protein conductances. They lead to a refractory period following a pulse during which the nerve is not excitable.  Thus, it is expected that nerve pulses traveling from opposite ends of a neuron will annihilate upon collision \cite{Tasaki1949}.

Due to its dissipative nature, the action potential in the Hodgkin-Huxley model should be accompanied by heat production. However, investigations of the initial heat resulted in the finding that, within experimental error, no such heat is released during the action potential \cite{Abbott1958, Howarth1968, Howarth1975, Ritchie1985}. A first phase of apparent heat release is followed by a second phase of heat absorption \footnote{In addition to the initial heat that is in phase with the voltage changes, there is a slow metabolic release of heat that is uncorrelated with the action potential \cite{Abbott1958}.}. The emission and reabsorption of the initial heat is exactly in phase with the observed voltage changes, and the integrated heat associated with the action potential is zero within experimental accuracy.  The data thus indicates that the action potential is an adiabatic (non-dissipative) phenomenon such as, e.g., a sound wave. This finding is in conflict with the HH model as acknowledged by Hodgkin (\cite{Hodgkin1964}, page 70).

The absence of net heat release combined with the experimental finding of mechanical dislocations during the action potential \cite{Iwasa1980a, Iwasa1980b} provided the motivation for attempts to explain the action potential as a propagating electromechanical pulse \cite{Heimburg2005c, Heimburg2007b, Andersen2009}. Due to the presence of lipid chain order transitions just below physiological temperature, the elastic constants of biomembranes display both a non-linear dependence on lateral density and dispersion \cite{Heimburg2005c}, i.e., frequency dependence of the sound velocity. This was shown to result in solitary mechanical waves with properties surprisingly similar to those of the action potential.  For instance, they propagate with a velocity of about 100 m/s (which is similar to the velocity of the action potential in the myelinated nerves of vertebrates) and display a reversible heat release as found in experiments. The change in nerve thickness associated with such solitary  waves is approximately 1\,nm, in agreement with the changes in membrane thickness associated with a phase change.  Given the known capacitive properties of lipid membranes, this thickness change and the associated decrease in membrane  area can produce voltage changes on the order of 100\,mV without any transverse flow of charge.  It was shown \cite{Heimburg2005c} that the thermodynamic properties of biological membranes support the propagation of solitary waves that display electric, thermal and mechanical changes consistent with those found in experiments.  In contrast to the Hodgkin-Huxley view, an electromechanical theory would not lead to annihilation of colliding pulses but rather to near-lossless penetration \cite{Lautrup2011}.  Given the difference between these predictions of the fate of colliding nerve pulses, it is important to investigate whether they annihilate or simply pass through each other. 

It is generally believed that the action potential is generated in the neuron at the axon hillock \cite{Kandel2000}.  Pulse propagation in the direction of the axon, the so-called orthodromic propagation, occurs from the soma towards the end of the synapses. However, in vertebrate and invertebrate nerve cells, the action potential can also be stimulated in regions remote from the axon hillock, e.g., ectopic sites located in axons or dendrites \cite{Pinault1995}.  Pulse propagation in the opposite direction, called antidromic propagation, can occur \cite{Pinault1995}. In fact, orthodromic and antidromic impulse propagation in neurons and other excitable tissues can be induced by electrical stimulation in the vicinity of the soma or in the distal part of the axon respectively. 

The simultaneous stimulation of orthodromic and antidromic pulses can lead to collision events.  As suggested above, such events can provide important information regarding the nature of signal transmission of information in neurons.  In spite of its relevance for understanding neuronal function and behavior, surprisingly little attention has been paid to such phenomena. The collision between two impulses was first investigated by Tasaki in 1949 \cite{Tasaki1949} using the motor fibers innervating the sartorious muscle of the toad. From his experiments, Tasaki concluded that the collision of two impulses result in their mutual annihilation.  Since this experiment was performed, little further work was done to confirm or to reject its finding.  This may be due in part to the fact that the outcome of Tasaki's experiment is in agreement with the predictions of the HH model \cite{Hodgkin1952b}. The importance of further investigation is emphasised by the fact that collision experiments, supplemented by the assumption that impulses always annihilate each other, are often used to identify axonal destinations of single cells in the central nervous system \cite{Fuller1976, Kimura1976, Murray1994}.
 
In the current work we report on collision experiments using the ventral cords of earthworm \emph{Lumbricus terrestris} and the abdominal ventral cord of lobster \textit{Homarus americanus} and show that the collision of two impulses generated simultaneously in orthodromic and antidromic directions does not result in their mutual annihilation.  Instead, they penetrate each other and emerge from the collision without material alterations of their shape or velocity.  The earthworm was chosen because of the properties of the median giant fibers (MGF) and because of the possibility of making simultaneous orthodromic and antidromic stimulation \cite{Mill1982}.  The electrotonic connections of the synapses and the neuronal syncytia permit the bidirectional propagation of the action potential along the array of giant neurons that form the MGF \cite{Gunther1975}. A similar situation is found in the median giant axons of the ventral cord of lobster \cite{Govind1976}. We compare these findings with simulations of the action potential as suggested by the electromechanical soliton theory.

\section{Materials and Methods}

\emph{Materials.} Earthworms (\textit{Lumbricus terrestris}) were obtained from a local supplier. We used an earthworm saline solution adapted from Drewes et al.  \cite{Drewes1974} consisting of 75 mM NaCl, 4 mM KCl, 2 mM CaCl, 1 mM MgCl, 10 mM Tris and 23 mM Glucose, adjusted to pH 7.4 with 8 mmol/l HCl. All the chemicals used in the preparation were purchased from Sigma-Aldrich.

Lobsters (\textit{Homarus americanus}) were obtained from a local supplier. We used a lobster saline solution adapted from \cite{Evans1976a} with the following composition; 462 mM NaCl, 10 mM KCl, 25 mM CaCl2, 8 mM MgCl2, 10 mM Tris and 11 mM Glucose, adjusted to pH 7.4 with NaOH.

\emph{Hardware and Software.} The Powerlab 26T data acquisition hardware was purchased from AD Instrument Europe (Oxford, UK). The instrument contains an internal bio-amplifier that allows recording small electrical potential on the order of microvolts. The bio-amplifier contains two recording channels (further description see AD instruments webpage). The Labchart software from AD Instruments was used to control the PowerLab 26T sending the stimulation and recording the signals coming from the ventral cord. 

\begin{figure}
    \centering
    \includegraphics[width=80mm]{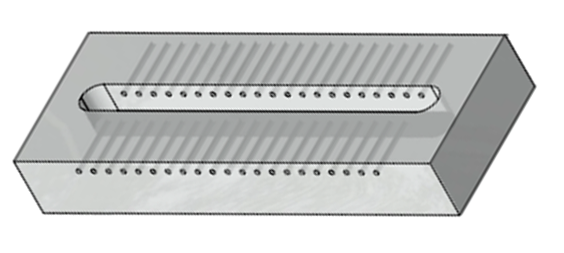}
    \caption{The design of the recording chamber. The nerve is placed on top of 21 electrodes located slightly above an aqueous buffer. The chamber is closed with a lid to avoid drying of the nerve. The electrodes at the end are used for stimulation, while the center electrodes are used for recording the signal.}
    \label{Figure1}
\end{figure}

\emph{Nerve chamber.} The self-built nerve chamber is composed of an array of 21 stainless steal electrodes in a longitudinal cavity covered by a lid in order to protect the nerve once extracted. The lid also allows maintenance of a saturated water vapor atmosphere in order to keep the moisture in the ventral cord. The nerve chamber is a 7 $\times$ 2.5 cm block of 1 cm height made on Plexiglas that contains a longitudinal channel of 6 cm length (depth 0.5 cm depth and width 0.5 cm). In the longitudinal aperture, an array of 21 perforations was created to place stainless steel electrodes. The array was located about 0.25 cm from the top of the chamber. The distance between consecutive electrodes is 0.25 cm. The stainless steel electrodes have a length of about 3.4 cm and a diameter of 0.5 mm and were fixed in the perforation along the chamber by using Reprorubber Thin Pur by Flexbar (Islandia, NY).  A scheme of the nerve chamber is shown on Fig. \ref{Figure1}.


\begin{figure}
    \centering
    \includegraphics[width=80mm]{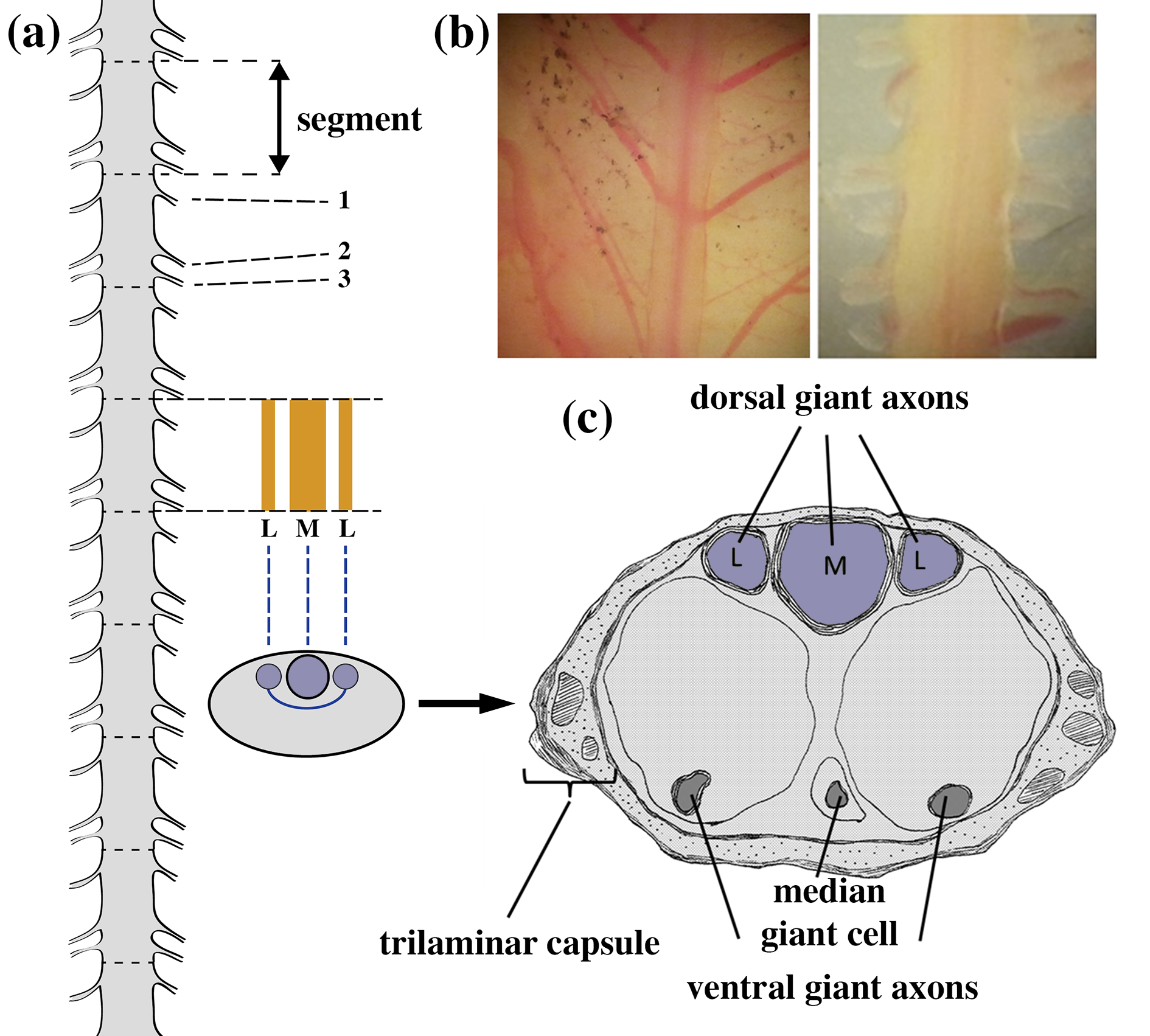}
    \caption{ (a) schematic representation of an earthworm ventral cord with the segments and 3 pairs of roots, (b) ventral cord with and without muscular tissue, (c) internal structure of the ventral cord redrawn from \cite{Coggeshall1965}.  Median and lateral giant fibers are marked with M and L.}
    \label{Figure2}
\end{figure}

\emph{\textbf{Nerve preparation of earthworm.}} The earthworms (\emph{Lumbricus terrestris}) were anesthetized by immersing them in a solution of  10\% ethanol in tap water. The earthworm was left between 5 and 10 min in the anesthetic solution depending on its size. Once removed from the anesthetic solution, the earthworm was washed with tap water to remove remains of the anesthetic solution and fixed longitudinally in a dissecting pan using pins. The earthworm was pinned laterally with the ventral side facing the dissection pan. A small incision in the dorsal side was made by using an scalpel or small scissors. Subsequently, the incision was elongated along the entire length of the earthworm body. Using micro-scissors with a straight blade, we cut each septum to liberate the internal organs and pin down the loose skin with muscular tissue. Using curved micro-scissors, we removed the crop, gizzard and intestine and the first 20 segments of the ventral cord including the brain. After this step we cleaned the preparation with the saline solution leaving the ventral cord and median ventral blood vessel exposed. In the final step we cut each segment below the ventral cord taking care to avoid damaging the sample. The blood vessels were removed before extracting the ventral cord.  The extraction and all experiments were performed at room temperature ($\sim$\,22$^\circ$C. A scheme of the internal and external structure of the ventral cord is shown in Fig.\ref{Figure2}.


\begin{figure}
    \centering
    \includegraphics[width=80mm]{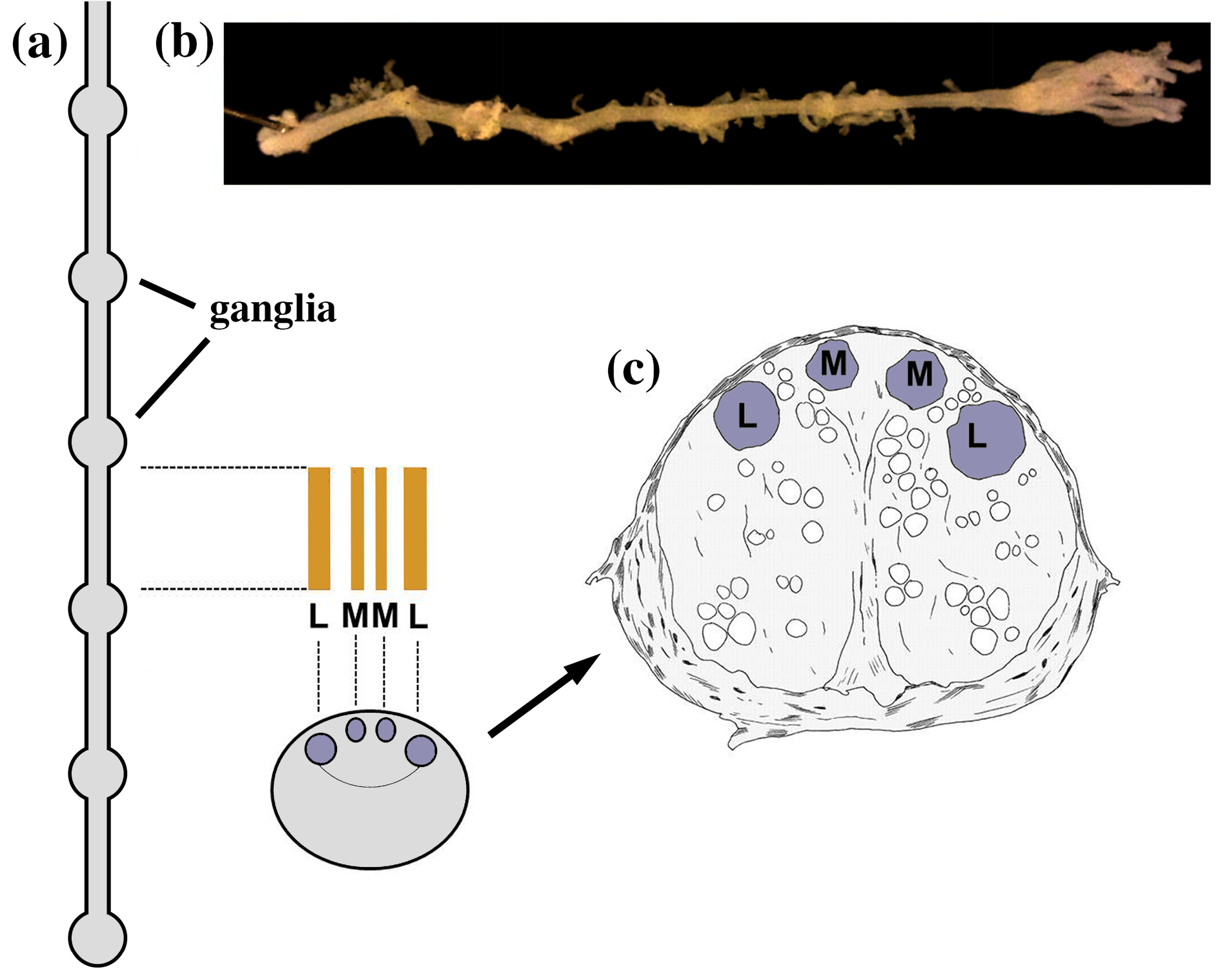}
    \caption{ (a) schematic representation of a lobster ventral cord at abdominal (tail) site with six ganglia, (b) abdominal ventral cord extracted from the lobster tail, (c) internal structure of the ventral cord showing four giant axons.  }
    \label{Figure2b}
\end{figure}

Electrical stimulation can be used to excite the median and lateral giant fibers (LGF) of intact anesthetized earthworms \cite{Kladt2010}. The signals coming from the small giant axons of the ventral side of the ventral cord cannot be detected using our setup. The signal from the three small giant axons require higher stimulation voltages and probably higher amplification. Note that even in the intact earthworm the signals from the median and lateral axons can be detected in external recordings. This is not the case for the small giant axons \cite{Kladt2010}.

However, earthworms can move during the experiment after several electrical stimulations.  Additionally, an excessive amount of anesthetic preventing movements during the experiment will affect the excitation properties of the ventral cord as well as the propagation velocity. For this reason we performed experiments on the extracted ventral cord.

The ventral cord from Lumbricus terrestris was used freshly after extraction and special care was taken to remove any remaining tissue that did not  belong to the ventral cord itself. The ventral cord was left in the Ringer solution for about 30 min,\ to relax.  We proceeded to the best step after the ventral cord had reached a stable size.  The ventral cord is very flexible and can be stretched without damaging its internal structure as a consequence of the trilaminar layer that protect the neurons filling the inside. After the equilibration period, the ventral cord was placed in the nerve chamber over the electrode array and a few microliters of the Ringer solution were deposited at the bottom of the chamber. The nerve chamber was closed with a glass lid, allowing an atmosphere of saturated water vapor to accumulate.  This prevented the loss of moisture by the ventral cord and the subsequent death of the nerve for several hours.   The ventral cord was placed with the ventral side facing the electrode array, and the preparation was ready for the collision experiment.  Two pairs of stimulation electrodes were placed close to the two ends of the extracted ventral cord.  A single pair of recording electrodes was placed at about 1/3 of the distance between the stimulation sites. As an initial step, we determined the voltage threshold for generation of an action potential propagating orthodromically.  We followed the conventional protocol by increasing the voltage in small intervals. The same protocol was followed by stimulating the ventral cord from the tail-side to generate an action potential propagating antidromically.  In all the experiments we found that slightly higher voltages were needed to initiate antidromic action potential propagation. At voltages higher than the voltage threshold the spike was stable and unchanged in shape and position. The observed differences in threshold voltage can be due to variations in diameter of the giant neurons along the MGF. The MGF becomes smaller in diameter towards the posterior end of the earthworm, and the LGF becomes smaller towards the anterior end of the animal \cite{Mill1982}. According to Coggeshall \cite{Coggeshall1965} the diameter of lateral giant axons ranges between 4 $ \mu $m in the anterior regions and 50  $ \mu $m in the posterior regions of the nerve cord while the diameter of the median giant fiber in the posterior end is of the order of 100 $ \mu $m.  In order to have a relatively uniform diameter in the median giant axons, we used a fragment of the ventral cord starting at segment 20 with a total length of about 4 to 6 cm. Because the two LGF are physically connected and fire in a synchronous way, we used only the MGF for the collision experiment \cite{Kladt2010}.  We should note that voltage values higher that those used for the antidromic action potential for the MGF will generate an antidromic action potential for the LGF, which we wanted to avoid. The median giant axon has a larger diameter than the lateral giant axon over the full length of the ventral cord fragment used in our experiments. This results in faster signal propagation in the median giant axon \cite{Lagerspetz1967} and makes it possible to distinguish the action potentials from the median and the lateral giant axons. In all cases we verified  for orthodromic and antidromic propagation that at higher voltages we could get a second signal with a bigger latency (corresponding to the LGF) ruling out any uncertainty in the spike identification.

\emph{\textbf{Nerve preparation of lobster.}} The Lobster (Homarus americanus) was anesthetized by keeping the animal in the freezer for about 30 minutes. Once removed from the freezer the animal was placed on the dissecting table and the head was severed in order to remove the brain. In a second step, a cut was made at the onset of the abdomen in order to separate the tail. The abdominal part of the ventral cord can be extracted by cutting both laterals of the ventral side of the animal and removing the soft shell. The abdominal ventral cord is attached to the soft shell and is easily removed with tweezers after cutting the nerves branching from the six ganglia. The ventral cord contains four giant axons. Two median giant axons that run, as a single neuron, all the way through the abdominal ventral cord and two lateral giant axons that are formed by 6 neurons connected at the level of each ganglia \cite{Johnson1924}. In the abdominal part, the lateral giant axons display a larger diameter than the median giant axons \cite{Govind1976}. They are excited at lowest stimulation voltage. A cross-section of the ventral cord is shown in Fig. \ref{Figure2b}. The giant axons of the ventral cord of lobster are considered non-myelinated \cite{Hartline2007}.

\section{Results}
\begin{figure}[soliton]
    \centering
    \includegraphics[width=80mm, height=80mm]{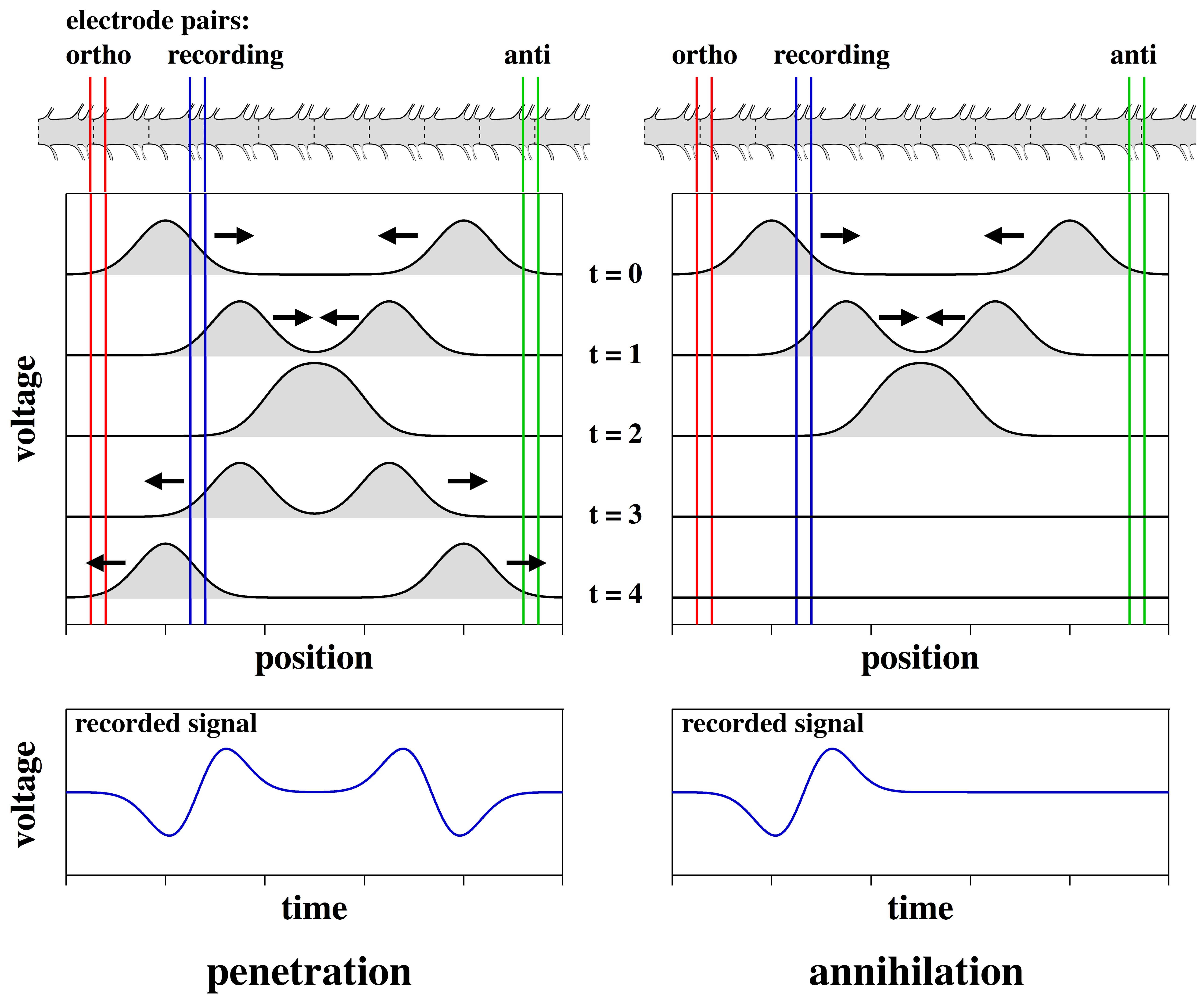}
    \caption{A schematic representation of the different steps of the collision experiment at times $t_0$, $t_1$, $t_2$ and $t_3$.  The action potentials (AP) are generated at $t_0$ by simultaneous stimulation in both ends of the ventral cord; the orthodromic AP reaches the recording electrodes at $t_1$; the APs collide at $t_2$; the antidromic AP reaches the recording electrodes at $t_3$.}
     \label{Figure3}
\end{figure}

A schematic description of key steps in the collision experiment is shown in Fig. \ref{Figure3}. We stimulate the axon with two pairs of electrodes at the two ends of the nerve (shown in red and green). After about 1/3 of the total length of the axon in orthodromic direction Two recording electrodes (shown in blue) are located at about 1/3 of the total length of the axon in the orthodromic direction. Since the difference in potential between these two electrodes is recorded, the resulting signal is approximately the first derivative of the true pulse shape. If two pulses are generated simultaneously at opposite ends of the nerve, the orthodromic pulse is recorded by these electrodes before the collision of the pulses while the antidromic pulse is detected after the collision. If the two pulses penetrate each other, the recorded signal will show both the orthodromic pulse and the subsequent antidromic pulse (see Fig. \ref{Figure3}, bottom left).  If the two pulses annihilate, only the initial orthodromic pulse will be recorded (see Fig. \ref{Figure3}, bottom right). The results for simultaneous stimulation at both ends can be compared with experiments in which only the orthodromic or only the antidromic pulse was stimulated.  Although the schematic drawing in Fig. \ref{Figure3} suggests that orthodromic and antidromic pulses have the same shape, their shapes can differ in real nerves because the thickness of the axon is not constant along its full length.  

\subsection{Theory}
Biological membranes display lipid chain melting transitions slightly below body temperature. In these transitions, the lateral compressibility of the membrane changes as a non-linear function of the lateral mass density. The compression modulus is also a function of frequency.   These two facts lead to the possibility of propagating mechanical solitons (or solitary pulses) \cite{Heimburg2005c}.
The mathematical expression for a propagating soliton in such a membrane cylinder is given by
\begin{eqnarray}\label{eq:01}
\frac{\partial^2}{\partial t^2} \Delta \rho & = & \frac{\partial}{\partial x} \left[
\left( c_0^2 + p \Delta \rho + q (\Delta \rho)^2 \right) \frac{\partial}{\partial x}
\Delta \rho \right] \nonumber \\
{} & {} &   - h \frac{\partial^4}{\partial x^4} \Delta \rho \ .
\end{eqnarray}
where $x$ is the spatial coordinate along the membrane cylinder and $t$ is time. Here, we use parameters appropriate for dipalmitoyl phosphatidylcholine (DPPC) membranes at 45 $^\circ$C as given in \cite{Heimburg2005c}. The density variation, $\Delta \rho = \rho - \rho_0$, is the difference between the lateral mass density of the membrane and its empirical equilibrium value of $\rho_0 = 4.035 \times 10^{-3}$\,g/m$^2$. The low frequency sound velocity is $c_0 = 176.6$\,m/s.  The coefficients $p$ and $q$ were fitted  to measured values of the sound velocity as a function of density. For the simulations here $p = -16.6 c_0^2/\rho_0$ and $q = 79.5 c_0^2/(\rho_0)^2$ found in Ref.\,\cite{Heimburg2005c}.   If the membrane is slightly above the melting transition of the lipid chains, it is to be expected that $p<0$ and $q>0$. The dispersion coefficient, $h$, must be positive. The above equation possesses exponentially localized solutions of a fixed shape which propagate with an arbitrary constant velocity, $v$, that is smaller than $c_0$ and larger than a minimum limiting velocity $v_{\rm min}$.  Eq.\,\ref{eq:02} possesses analytic solutions given in \cite{Lautrup2011}. 

The pulse amplitude reaches a maximum amplitude of
\begin{equation}
\label{eq:02}
\Delta \rho_{\rm max}=\frac{|p|}{q}
\end{equation}
as the velocity approaches the limiting value \cite{Heimburg2005c} of 
\begin{equation}
\label{eq:03}
\Delta v_{\rm min}=\sqrt{c_0^2-\frac{p^2}{6q}} \, .
\end{equation}
Thus, different velocities are associated with different pulse amplitudes and energies. For synthetic DPPC large unilamellar vesicle membranes slightly above their melting temperature, the minimum pulse velocity is $v_{\rm min}= 0.65 c_0$ and the maximum amplitude change is  $\Delta \rho_{\rm max}/\rho_0=0.209$.  This corresponds to passing from the liquid to the solid phase of the membrane.

\begin{figure}[htb!]
    \centering
    \includegraphics[width=60mm]{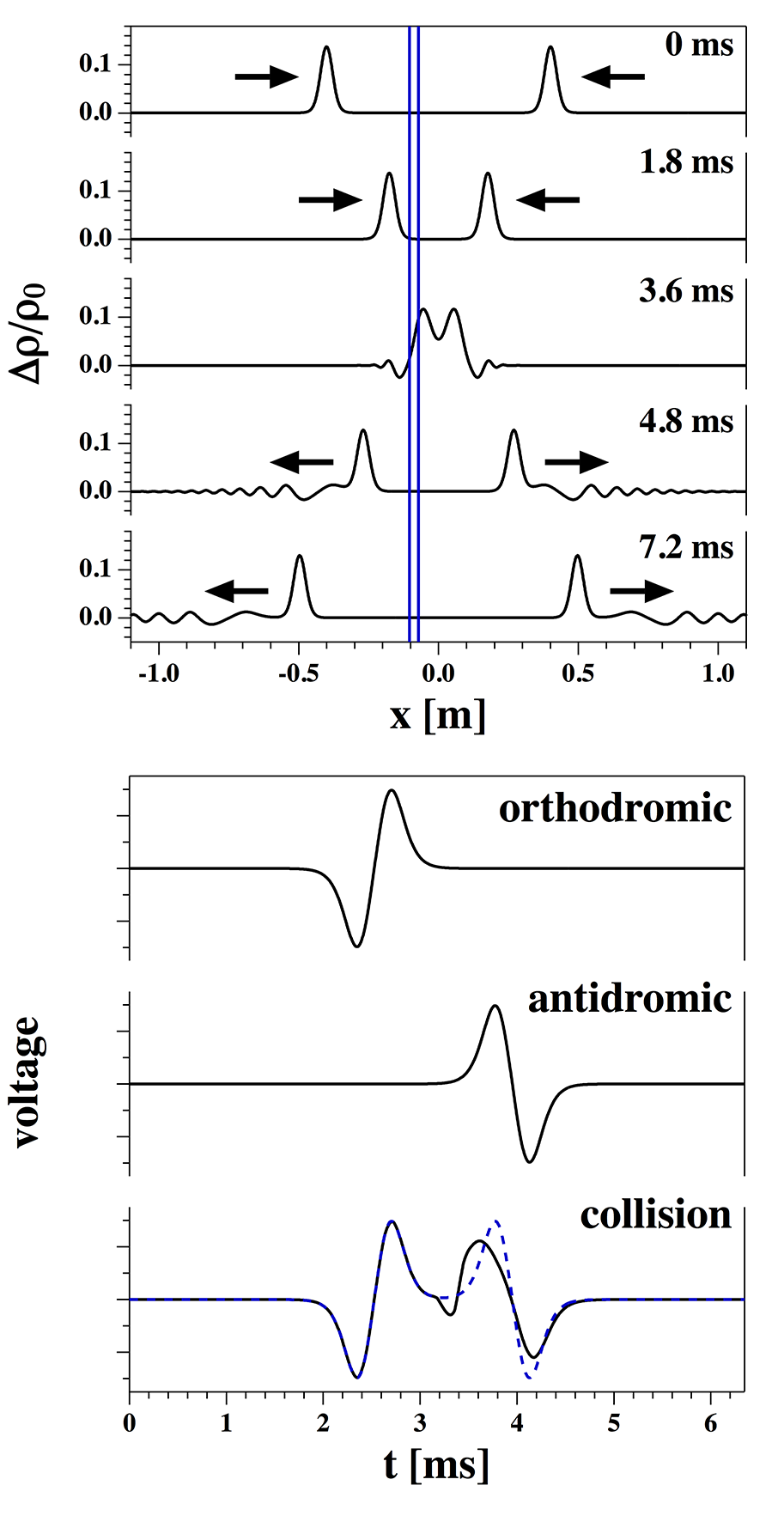}
    \caption{Top: The collision of two pulses in the soliton theory of nerves for v=0.7c$_0$. Parameters are given in the text. After collision, the shape of the solitary pulses is virtually unchanged. The two blue lines indicate the positions of two hypothetical recording electrodes with a distance of 16\,mm. Bottom: The calculated voltage difference between the two electrodes is shown in the top panel. The top trace shows the single orthodromic AP, the center trace shows the antidromic AP, and the bottom trace (solid) shows the recording of the two colliding pulses. The dashed blue line is the sum of the ortho- and antidromic without a collision. It is added as a guide to the eye. }
    \label{FigureT}
\end{figure}

We have solved eq.\ref{eq:01} numerically using the above parameters for DPPC and a velocity of $v=0.7 c_0$. The solid lipid phase has a maximum density that is 24.6\% higher than that of the liquid state. In order to prevent densities higher than that of the solid lipid phase during pulse collision, we have introduced a soft barrier at $\Delta \rho/\rho_0 \approx 0.25$ (see \cite{Lautrup2011} for details). The results are shown in Fig.\ref{FigureT}. The top panel shows the pulses propagating before and after the collision at five different times $t$.   The collision process leads to some dissipation of energy in the form of small amplitude noise that propagates with the speed of sound $c_0$ (i.e., faster than the velocity of the solitary pulse).  The shape, velocity and energy of the pulses are largely unaltered.  During pulse collision, the density changes of both pulses do not have to add as one intuitively might assume. Instead, one finds a broadened intermediate collision state, which is wider than the individual solitons. In the soliton theory collision obviously does not lead to annihilation of the colliding pulses. The fact that the individual pulses suffer a minor loss of energy during the collision merely indicates that we are considering solitary pulses rather than true solitons. The generation of small amplitude noise with very low energy content is mostly a consequence of not allowing the density change to exceed $\Delta \rho/\rho_0 = 0.25$.

Biomembranes can be regarded as charged capacitors \cite{Heimburg2012}. Voltage changes are directly related to the density changes. Assuming that $\Delta \rho$ is proportional to a change in voltage, we can determine the voltage signal recorded by two hypothetical electrodes that are placed as shown in Fig. \ref{Figure3}. These electrodes are shown as blue lines in Fig.\ref{FigureT} (top) and, in our simulation, they are separated by 1.6 cm (which is to be compared with the total pulse width of about 5 cm). Fig.\ref{FigureT} (bottom) shows the recording by these electrodes for a single pulse from the left, a single pulse from the right and for the calculated collision experiment. The dashed blue line is the sum of the two single pulses shown as a guide for the eye.  It is clear that the second pulse is distorted by the collision process as it is expected from the analysis in the top panel. 

The results of soliton theory described above can be compared with the well-known Hodgkin-Huxley model.  Originally designed to describe a squid axon containing sodium and potassium channel proteins, the differential equation for the Hodgkin-Huxley model is given by 
\begin{eqnarray}
\label{eq:04}
\frac{r}{2 R_i}\frac{\partial^2}{\partial x^2}V&=&C_m\frac{\partial}{\partial t}V+ g_K(V,t)\left(V-E_K\right) \\
&&+  g_{Na}(V,t)\left(V-E_{Na}\right)\nonumber
\end{eqnarray}
where the transmembrane voltage $V$ is the observable (instead of $\Delta \rho$), $r$ is the radius of the axon, $R_i$ is the resistance of the cytoplasm in the axon,  $C_m$ is the membrane capacitance and $E_K$ and $E_{Na}$ are the Nernst potentials of potassium and sodium reflecting the differences in ion concentrations inside and outside of the neuron. The conductances of potassium and sodium ions, $g_K(V,t)$ and $g_{Na}(V,t)$, are complicated functions of voltage and time. If additional channel proteins are present, more conductance terms must be added to eq.\,\ref{eq:04}. Eq. \ref{eq:04} has a structure similar to the wave equation \ref{eq:01}.  However, no general theory exists for the conductances $g_i(V,t)$.  Their dependence on time and voltage must be determined empirically from voltage-clamp data \cite{Hodgkin1952b}.  This introduces many parameters into the above equation. Further, not all nerves are as simple as the squid axon, and they may contain more than just two channel proteins.  More terms containing the conductances of other proteins must be introduced, further increasing the number of parameters.  In the literature one finds models with up to 66 parameters \cite{Howells2012}. Since different nerves contain different ion channels, it is not generally possible to make a generic statement about the pulse collision process. However, on the basis of numerical simulations for the squid axon, it is generally believed that the Hodgkin-Huxley model results in the annihilation of colliding pulses.  Qualitatively, this is due to a refractory period introduced by time-dependent changes in protein structure during the nerve pulse that render the nerve unexcitable for a short period after the pulse. This will be discussed further in the Discussion section.

\subsection{Experimental results}
\subsubsection{Earthworm experiments: } 
After confirming that we could stimulate the ventral cord from both ends independently, we performed a collision experiment by simultaneously stimulating both ends of the ventral cord. We proceeded by increasing the stimulating voltage in small intervals up to the values necessary to generate action potentials at both ends of the ventral cord fragment simultaneously. The recording electrodes are located closer to the site where the orthodromic pulse is generated.  It is therefore necessary for the antidromic pulse to pass through the orthodromic pulse before it can reach the recording electrodes (cf. Fig.\ref{Figure3}). Thus, if both the orthodromic and the antidromic pulse can be recorded, the two pulses must have passed through each other. If only the orthodromic pulse (but not the antidromic pulse) can be recorded, this is evidence for pulse annihilation. 
\begin{figure}[t]
    \centering
    \includegraphics[width=80mm]{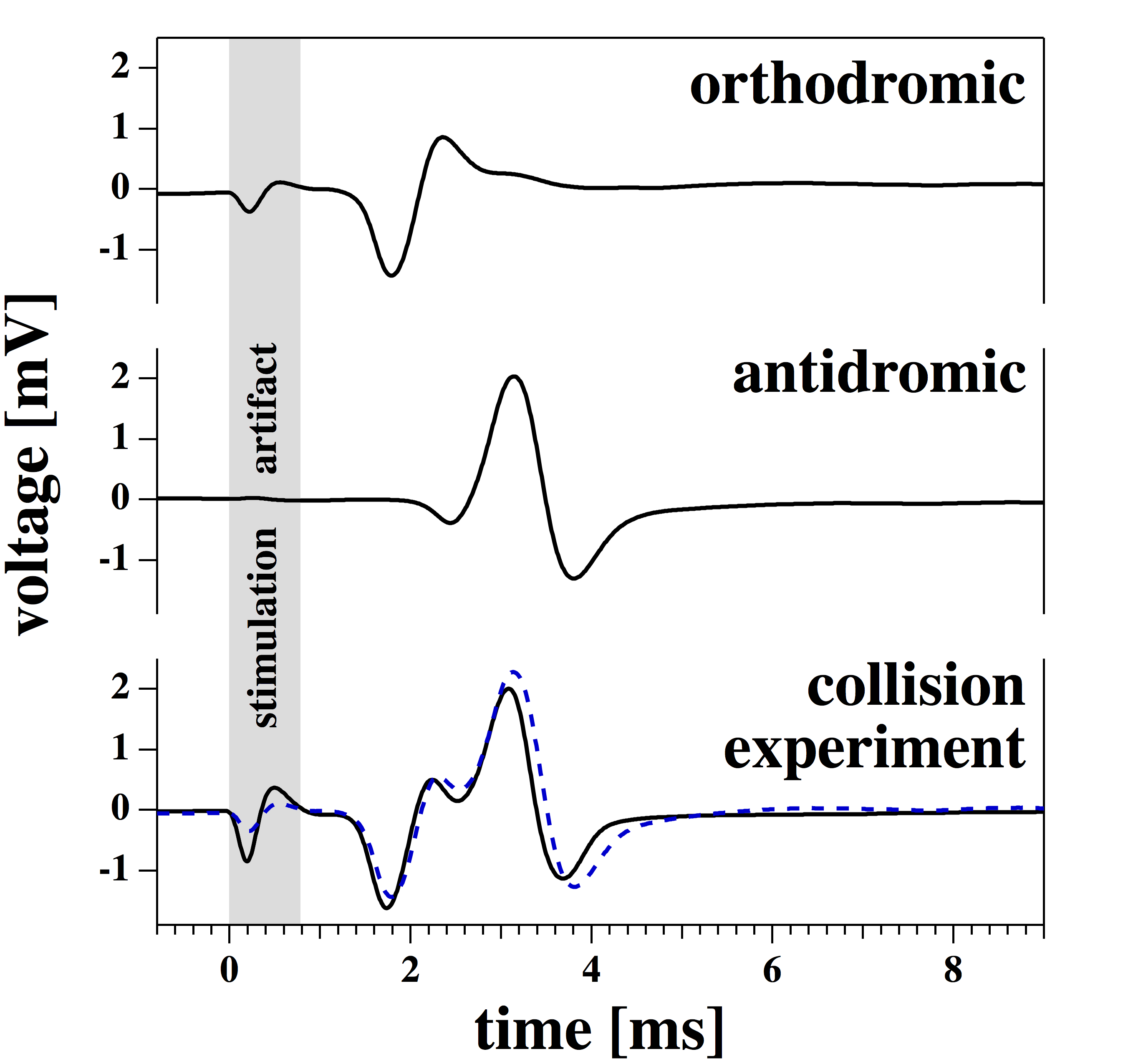}
    \caption{ Example of the pulse collision experiment in the ventral cord of earthworm (sample \#7 in Table \ref{Table1}). Top: Action potential propagating orthodromically after stimulation at the top end of the nerve. Center: Action potential propagating antidromically after stimulation at the bottom end of the nerve. Bottom: Both action potentials generated by simultaneously stimulation at both ends (solid line). For comparison and a guide for the eye, the dashed line represents the sum of the individual pulses. This signal is similar to the observed trace.  The region shaded in grey shows the stimulation artifact. }
    \label{Figure4}
\end{figure}

A representative result is shown in Fig. \ref{Figure4}. The top two traces show the orthodromic and antidromic pulses after individual stimulation. The antidromic signal arrives at the electrodes about 1.5 ms later than the orthodromic pulses. This interval is comparable to the width of the pulses. Therefore, one can recognize both pulses as separate events. The bottom trace shows the experiment where both orthodromic and antidromic pulse were generated simultaneously. We find that both pulses can be recorded and that they are unchanged in shape. As a guide to the eye, we show the sum of the individual orthodromic and antidromic pulses in the absence of a collision (dashed blue line). This signal is very similar to that recorded in the collision experiment indicating that pulse collision does not generate much distortion of the signal. This experiment was reproduced in at least 30 different worm axons, and we always found pulse penetration. We observed infrequent events (less than 15 \%) in which we recorded only the orthodromic pulse. This typically happened when the axon was moved such that the stimulation electrodes were close to the extreme ends of the axon. In all of these cases, relative movement of the same axon with respect to the electrodes reestablished pulse penetration.

\begin{table}[!ht]
\centering
\begin{tabular}{|c|cc|cc|}
\hline
&\multicolumn{2}{c|}{single} & \multicolumn{2}{c|}{simultaneous}  	   \\
   sample   & orthodromic  & antidromic  & orthodromic  & antidromic \\ 
   \hline
    1             & 6.60 (5.43)    & 2.78 (2.32)    & 6.25 (5.01)     & 2.68 (2.23)    \\
    2             & 6.29 (5.52)    & 3.54 (3.03)     & 6.12 (5.12)   & 3.16 (2.78)    \\
    3             & 8.12 (7.76)   & 6.23 (5.39)     & 7.99 (7.03)    & 5.83 (5.61)    \\
    4             & 8.46 (7.06)   & 6.97 (5.69)     & 8.23 (6.83)    & 6.71(5.41)    \\
    5             & 7.89 (6.63)   & 5.43 (4.82)    & 7.77 (6.48)    & 5.44 (4.91)  \\
    6             & 7.29 (6.65)   & 5.45 (4.71)    & 7.37 (6.77)    & 5.58 (4.91)   \\
    7             & 9.67 (7.81)   & 7.50 (6.66)    & 9.57 (7.77)    & 7.13 (6.69) \\
    8             & 7.76 (6.46)   & 3.98 (3.28)      & 7.45 (6.21)     & 4.04 (3.38)  \\
    9             & 7.51 (7.01)    & 4.02 (3.59)   & 7.79 (7.35)      & 4.12 (3.79)   \\
    10           & 8.37 (7.36)    & 6.13 (5.27)    & 8.18 (7.16)      & 6.01 (5.10)  \\
   \hline
    \end{tabular}
    \caption {Conduction velocity estimates in m/s from ten different collision experiments 0n the ventral cord of earthworm. All measurements were carried out at 22\,$ \pm$\,1\,$^{\circ}$C. The convention is to calculate the velocities by using the first extremum in each pulse recording. Values in brackets correspond to velocities calculated for the nodal point in each pulse that corresponds to the pulse maximum. The recordings belonging to sample \# 7 are shown in Fig. \ref{Figure4}.}
     \label{Table1}
\end{table}
\begin{figure}[htb!]
    \centering
    \includegraphics[width=6.0cm]{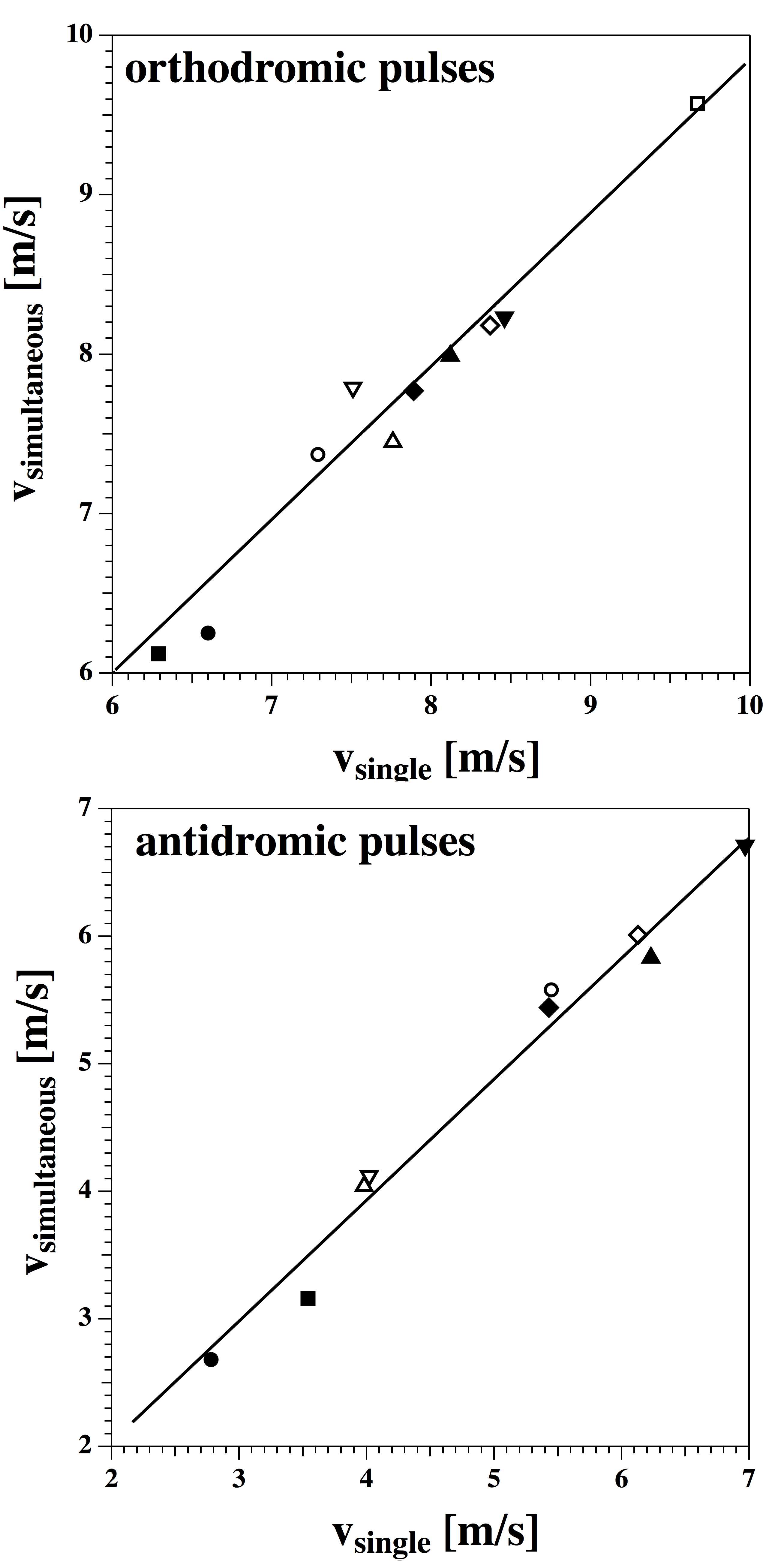}
    \caption{Pulse velocities obtained in the collision experiment in earthworm (simultaneous stimulation) versus the velocities of the single stimulation (using the first extremum of each pulse recording). Top: Orthodromic pulses. Bottom: Antidromic pulses. Same symbols in both panels indicate identical nerve preparation. The open square corresponds to the traces in Fig. \ref{Figure4} (sample \#7). Experimental temperature was  22 $ \pm $ 1  $^{\circ}$C.}
    \label{Figure5}
\end{figure}

In Table \ref{Table1} we report the results of a selection of 10 different samples (out of the 30 different nerves) for which there was little or no overlap between the orthodromic and antidromic pulses.  In these cases the velocities of the individual pulses could be determined easily. The convention is to determine the velocities from the first extremum in each pulse recording. For completeness, we also give the velocities for the nodal point in each trace corresponding to the pulse maximum (values in brackets). These values are somewhat smaller  but are also comparable for single pulses and pulses in the collision experiment. We show both the velocities of the orthodromic and the antidromic pulses in the case of single and of simultaneous stimulation. In general, the velocities of the antidromic pulses are lower.  This lower velocity could be a result of diameter changes along the median giant axon. Pulse velocities range between 2.8 and 9.7\,m/s. The earthworm MGF and LGF axon are considered myelinated (with varying degrees of myelin packing). The conduction speed is a few fold higher than that of non-myelinated fibers of the same diameter \cite{Hartline2007}.  Since the temporal width of the pulses is about 2\,ms, this corresponds to a lateral extension of the pulse of 4-17\,mm. Thus, the pulse width is larger than the distance between the electrodes. It is also larger than the average neuron in the segmented giant axons which has a length of about 1-1.5\,mm \cite{Gunther1975}.  In Fig. \ref{Figure5} we plot both orthodromic and antidromic pulse velocities in the collision experiment versus the velocities in the single stimulation experiment. Within experimental error, we find the the propagation velocities are unaltered by the collision. Similarly, as can be seen in Fig. \ref{Figure4}, the pulse shapes are unaltered by a collision. The pulse velocities are typically smaller in antidromic direction as compared to the orthodromic direction. We believe that this is due to the change in diameter of the fibers along the worm axis.

The interpretation of the collision experiments shown in Fig. \ref{Figure4} and Table \ref{Table1} rests on the assumption that in both orthodromic and antidromic direction the same fiber was stimulated. Early experiments from \cite{Kao1956, Kao1957} on the neural cord of earthworm show that the MGF displays a lower threshold voltage than the LGF. However, there is a finite possibility that in our collision experiments we stimulate the MGF in one direction and the LGF in the other direction. Under such circumstances, the action potentials would trivially pass by each other and never collide at all. As a consequence, this would lead to a misinterpretation of the experiment. 

\begin{figure}[t]
    \centering
    \includegraphics[width=80mm]{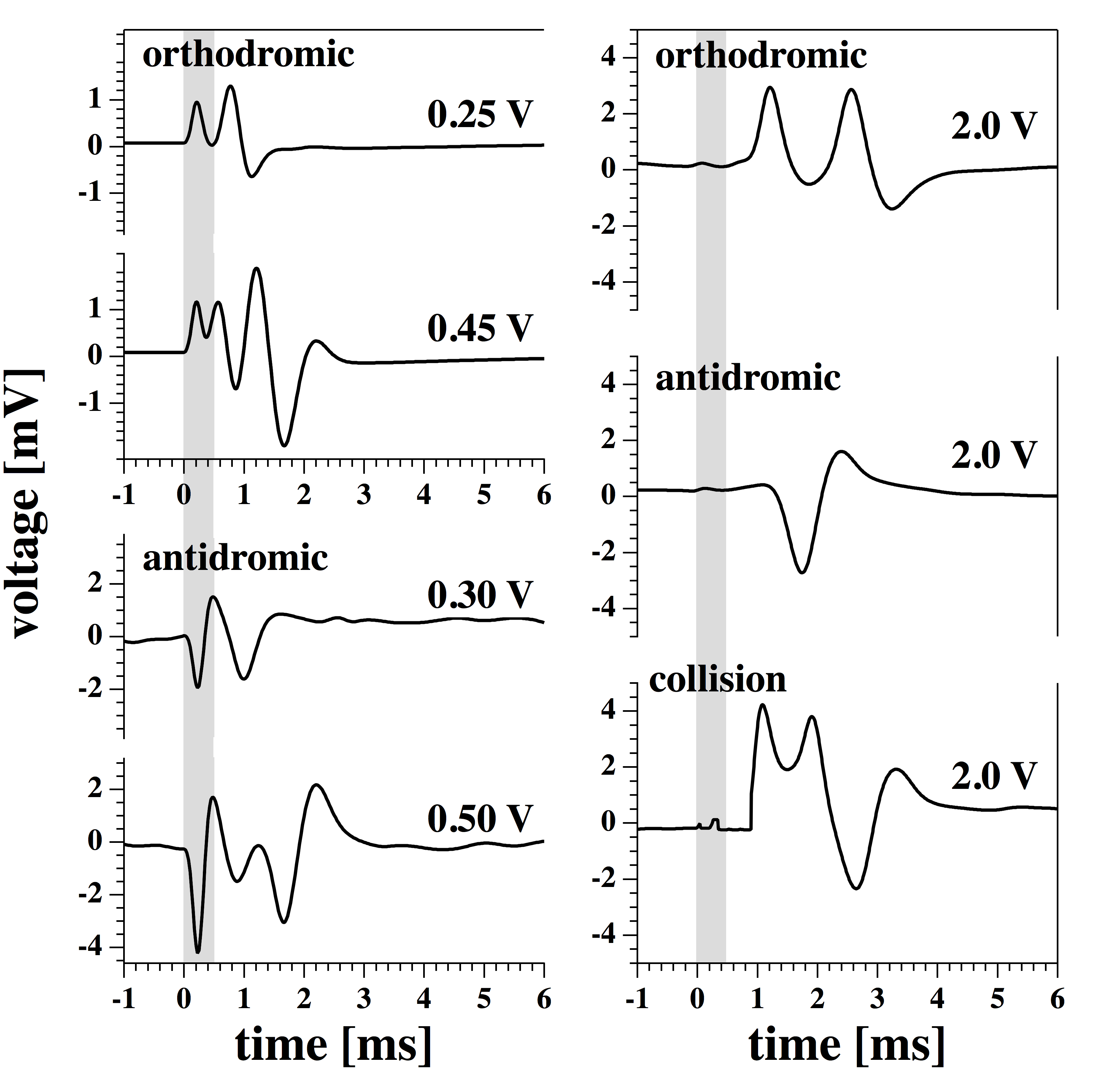}
    \caption{Simultaneous stimulation of both, MGF and LGF, in earthworm. Left, top: Single and double stimulation in orthodromic direction only. Left, bottom: Single and double stimulation in antidromic direction direction only. Right: Collision experiment with stimulation of both the MGF and LGF in orthodromic direct. In antidromic direction, only the MGF is stimulated. One can recognize that the antidromic signal is still present in the recording after collision. The grey-shaded regions mark the stimulation artifact. The left panels and the right panel were from different axons.}
    \label{Figure6}
\end{figure}

To rule out this possibility, we performed another set of experiments using double stimulation of both MGF and LGF. In Fig. \ref{Figure6} (left, top) we show the stimulation of the earthworm axon at two different stimulation voltages. At 0.25 V one observes only an action potential in one of the fibers. According to literature, this is likely to be the pulse in the MGF. At 0.45V, one sees both the pulse in the MGF and the LGF. Both voltages are directly above threshold for single and double stimulation. Thus, in order to stimulate both fibers nearly twice the stimulation voltage is required. A similar observation is made for the antidromic signal (Fig. \ref{Figure6} left, bottom). Here, too, one nearly needs twice the voltage to stimulate both pulses. Next we now performed an experiment in which both fibers were stimulated in orthodromic direction (Fig. \ref{Figure6}, right, top trace) and only one fiber was stimulated in antidromic direction (Fig. \ref{Figure6}, right, center trace)\footnote{Note that this is an experiment on a different axon as in the left hand panels. The shapes of the pulses and the respective stimulation voltages are different.}. After collision, one can still observe both action potentials in the orthodromic direction and the single action potential in antidromic direction. Independent of which fiber was stimulated in antidromic direction, it was unavoidable that it had collided with one of the two action potentials in orthodromic direction. This demonstrates that the antidromic pulse did not annihilate upon collision.

From Figs. \ref{Figure4} to \ref{Figure6} and the data in Table \ref{Table1}, we conclude that action potentials in the giant axons pass through each other without significant distortion.\\

\subsubsection{Experiments on giant axons from the abdominal part of the ventral cord of lobster: }
\begin{figure}[t]
    \centering
    \includegraphics[width=80mm]{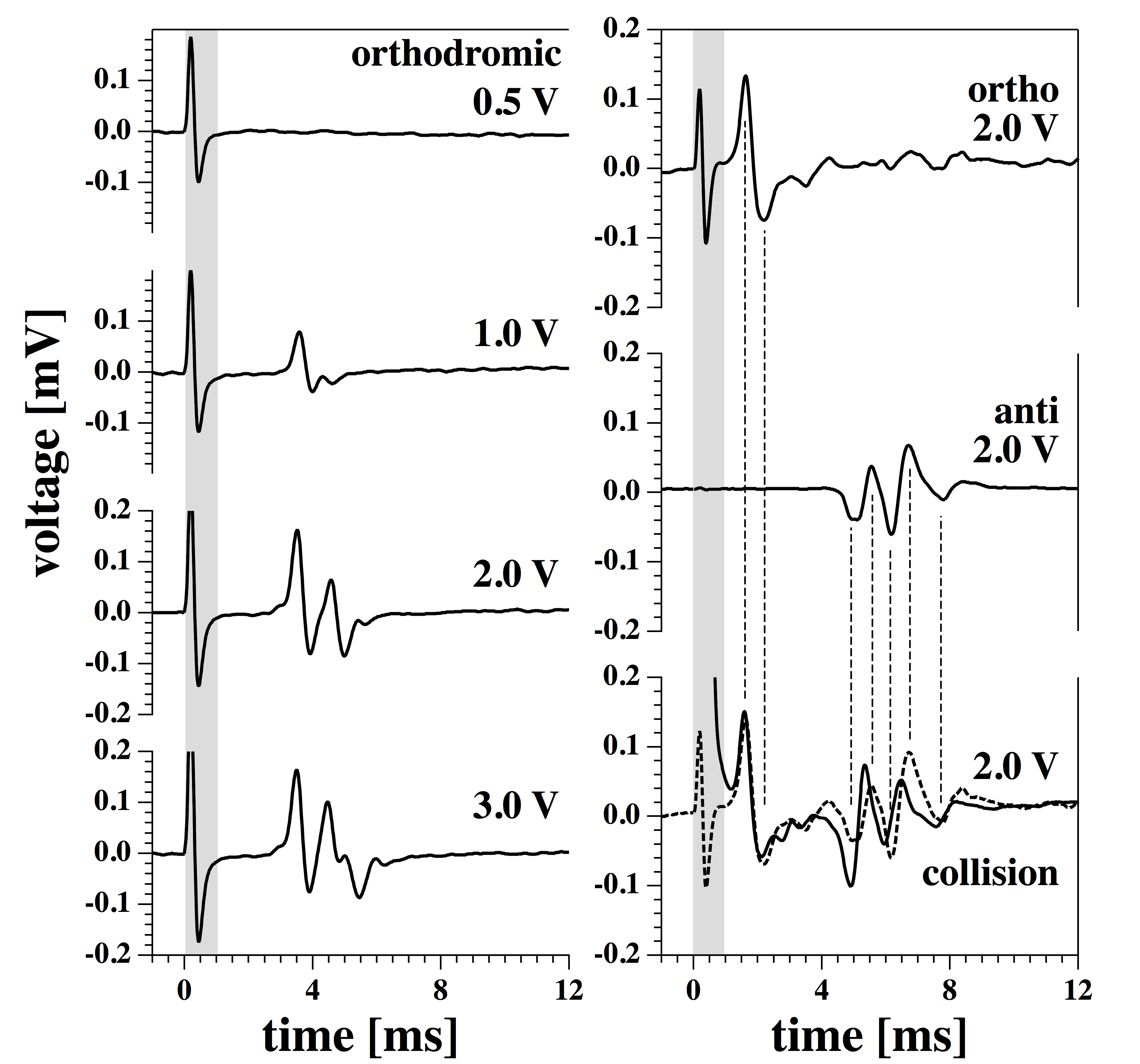}
    \caption{Left: Action potentials after stimulation in orthodromic direction show the successive generation of action potentials in the giant axons when increasing the stimulation voltage. Collision experiment in the abdominal part of the ventral cord of lobster at a stimulation voltage of 2V. Top: Stimulation in orthodromic direction only. Center: Stimulation in the antidromic direction only. Bottom: Collision experiment (solid line) compared with the sum of the top (orthodromic) and the center (antidromic) traces (dashed line). The two traces are virtually superimposable.}
    \label{Figure7}
\end{figure}

\begin{figure}[t]
    \centering
    \includegraphics[width=80mm]{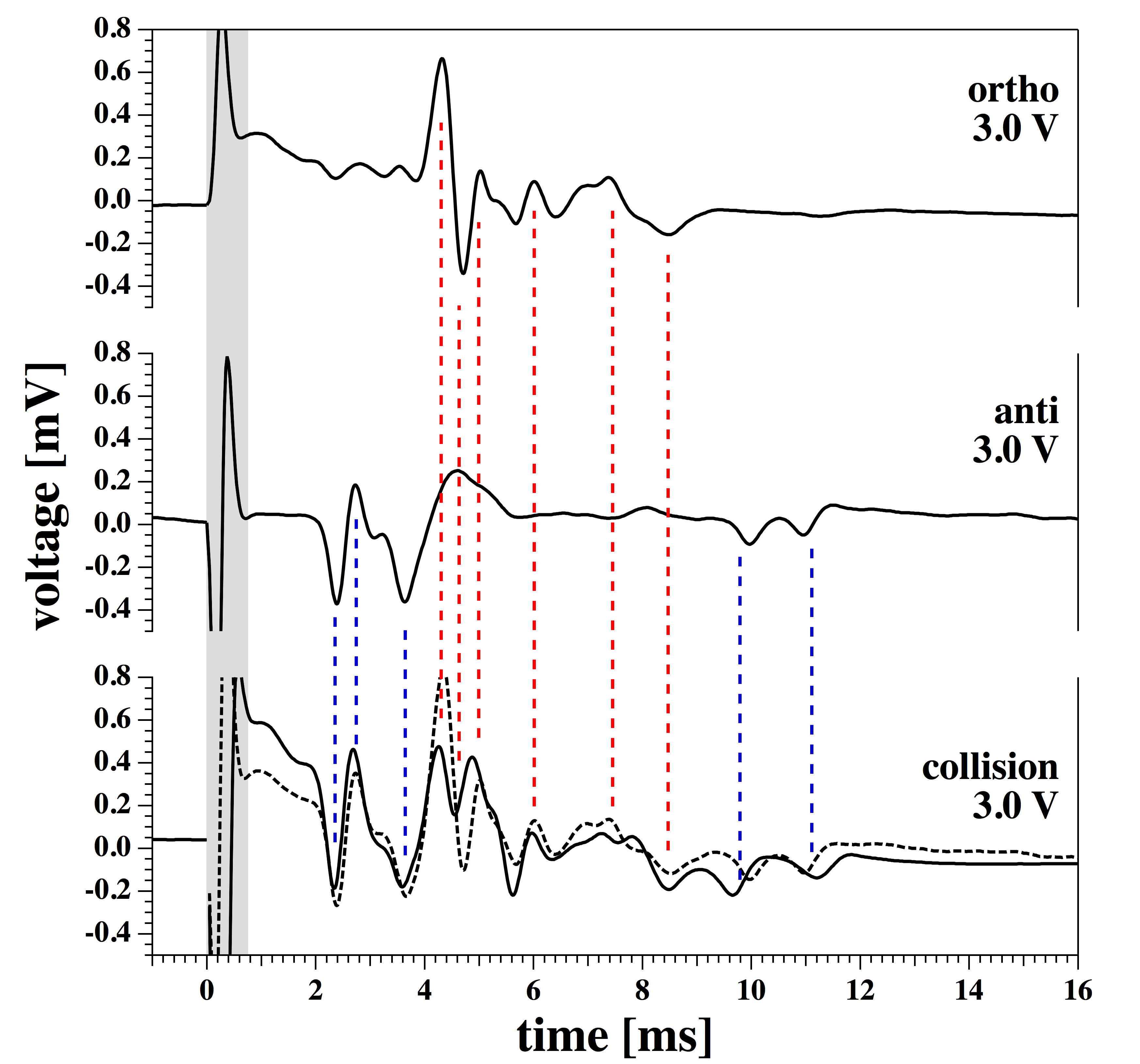}
    \caption{Same as in Fig. \ref{Figure7} (right) but with higher stimulation voltage (3V, different preparation). More axons are stimulated and all giant fibers are active. The bottom trace shows the collision experiment (solid line) compared with the sum of the top (orthodromic) and the center (antidromic) traces. Again, the two traces are virtually superimposable indicating that no annihilation of any of the signals took place.}
    \label{Figure8}
\end{figure}

In contrast to the earthworm, the ventral cord of lobster possesses two median and two lateral giant axons (Fig. \ref{Figure2b}). The median axon is not segmented as in the ventral cord of the earth worm. It has been described in the literature that in the abdominal part of the ventral cord the first (i.e., at lowest stimulus) and largest electrical signals correspond to the LGFs \cite{Govind1976}. The MGF pulse  (which displays a slower velocity than the LGF) appears as a next electrical signal upon increase in stimulus voltage. The small fibers in the ventral cord generate small signals and require high stimulation voltage. Fig. \ref{Figure7} (left) shows that an increasing number of giant axons is stimulated upon increase in voltage. In Fig. \ref{Figure7} (right) shows an experiment with one major orthodromic signal (top) and two antidromic (center) signals. Fig. \ref{Figure7} (right, bottom) shows the collision experiment. The dashed line is the sum of the orthodromic and antidromic pulses from the single side stimulation experiments. It can be seen the summed individual signals are nearly indentical to the signal in the collision experiment. It is most likely that the three signals in this experiment correspond to the lateral giant fibers of the ventral cord. Thus, one can conclude that pulses in one of the lateral fibers have passed through each other and did not annihilate. However, one cannot fully exclude the possibility that different neurons were stimulated in the the two directions, e.g., one signal in the LGF in one direction and two signals in the MGFs in the other direction. Under such conditions, pulses would actually never collide. For this reason, we repeated the experiment in the ventral cord of a different preparation at higher stimulation voltage (Fig. \ref{Figure8}). Now, more action potentials are excited in both orthodromic and antidromic direction (at least four in each direction.  The antidromic signal displays some signals with slow velocity that probably correspond to the MGF fibers. Thus, all giant fibers are stimulated. The bottom trace in Fig. \ref{Figure8}  shows the collision experiment. It shows that all signals in the collision experiment are conserved compared to the summed signals of orthodromic and antidromic stimulation. None of the signals was annihilated upon collision. We take this as convincing evidence that annihilation upon collision is not observed in the abdominal part of the ventral cord of lobster. This experiment was reproduced in eleven different preparations. Additionally, we repeated these experiments in other nerves from lobster including nerves from the legs and the connectives close to the lobster brain. In total, the above experiments were confirmed in thirty different nerve bundles from the walking legs, sixteen preparations of the thorax ventral cord and twelve samples from lobster connectives. Those results will be reported independently.


\section{Discussion}
We investigated the collision of action potentials in giant axons of the earthworm both experimentally and theoretically. Orthodromic and antidromic pulses were stimulated at both ends of the isolated axon. We showed in at least 30 independent nerve preparations that colliding action potentials pass through each other without significant perturbation. In less than 15\% of the preparations we found annihilation of pulses. In all of these cases, penetration could be reestablished by slight changes in the position of the axon on the electrodes. We believe that these cases reflect effects related to the extreme ends of the axon. We confirmed these findings in preparations from the ventral cord of lobster. When exciting all giant axons at large stimulation voltage at both ends of the nerve, all signals in orthodromic and antidromic direction were maintained after collision without major perturbation. No evidence for pulse annihilation was found. In nonlinear hydrodynamics simulations we further studied the penetration of pulses using the soliton theory  \cite{Heimburg2005c, Lautrup2011}.  As expected, this theory indicates that solitary pulses pass through each other with the production of minor amounts of small amplitude noise. This is consistent with our experimental finding. However, it is seemingly in conflict with expectations based on electrophysiological models such as the Hodgkin-Huxley model \cite{Hodgkin1952b} where a refractory period is expected to lead to pulse annihilation.

Indeed, it is widely believed that action potentials do annihilate upon collision. However, pulse annihilation is not well documented in the experimental literature. The most relevant report by I. Tasaki from 1949 \cite{Tasaki1949} discussed annihilation in myelinated nerve fibers in the Sartorius muscle of the toad. Tasaki reported pulse annihilation in this preparation. The analysis of the results involved saltatory conduction between the nodes of Ranvier in the myelinated nerve. To our knowledge, Tasaki's experiments never were reproduced. In 1982, Tasaki and Iwasa reported the mechanical response of colliding pulses in squid axons \cite{Tasaki1982d}. They found a slight modification of the mechanical pulse at the site of the the collision, but pulse annihilation was not examined. We have not succeeded in finding further original publications on pulse annihilation, and it is not certain that the common notion of the existence of pulse annihilation is well-rooted in experiment. However, within the context of the Hodgkin-Huxley model, it seems natural to expect pulse annihilation on theoretical grounds. The refractory period is a brief period after stimulation of an action potential during which the nerve is not excitable. It has been found in many nerves. Talo and Lagerspetz \cite{Talo1967} reported refractory periods of 1.2 -1.5\,ms around room temperature both for median lateral fiber and lateral giant fiber of earthworms.  Kladt et al. \cite{Kladt2010} reported refractory periods of 0.7-2.8 ms in intact earthworms.  These numbers are comparable to those found by us ($\approx$2\,ms, data are not shown). Thus, empirically short refractory periods exist in earthworms and many other nerves. As our experiments show, the existence of a refractory periods does not automatically imply the annihilation of colliding action potentials. It seems plausible to postulate that two colliding pulses annihilate because they are expected to enter into unexcitable regions of the neuron immediately after their collision. In the context of the HH-model, the existence of refractory periods is a consequence of relaxation processes in channel proteins after firing. The original model considers only sodium and potassium channels. However, neurons from other sources may contain many different Na- and K-channels as well as many other channel proteins such as calcium channels. Thus, one cannot easily generalize the properties of a particular neuron such as the squid axon for which the HH-model was designed. Since there exists no general theory for the voltage-dependent and temporal behavior of channel proteins, the properties of such proteins are typically parametrized from experiment. The model by Bostock and collaborators for the myelinated axons in humans contains 66 parameters describing 5 different channels that display different concentrations in different regions of the nerve \cite{Howells2012}. The Fitzhugh-Nagumo model \cite{Fitzhugh1955, Fitzhugh1961, Nagumo1962} is a simplification of the  Hogkin-Huxley model. It has been shown, that in this model (using only sodium and potassium channels) possesses parameter regimes in which pulses can penetrate \cite{Aslanidi1997}. Thus, it seems that the Hodgkin-Huxley model does not necessarily exclude the possibility of penetrating pulses. Interestingly, Tasaki dismissed the idea that the refractory period is responsible for pulse annihilation in his original publication from 1949 \cite{Tasaki1949}. He rather believed that during pulse collision the currents inside and outside of the nerve add up to zero such that the condition for regenerating the pulse is not met during collision.

The experiment by Tasaki \cite{Tasaki1949} on toad nerves indicates that there may be examples for pulse annihilation (even though reproducing this experiment would be helpful). However, we can falsify the general belief that annihilation must always occur due to the presence of a refractory period. Here, we have demonstrated penetration of pulses in myelinated (earthworm) and non-myelinated (lobster) giant axons.

The notion of penetrating pulses is not consistent with the Hodgkin-Huxley model if there is a refractory period (such as earthworm axons).  It is, however, in agreement with the assumption of the existence of mechanical pulses in nerves. Mechanical dislocations in various nerves have been experimentally confirmed in squid axons, and nerves from crab, garfish, \cite{Iwasa1980a, Iwasa1980b,  Tasaki1980, Tasaki1982a, Tasaki1982b, Tasaki1982d, Tasaki1982e,Tasaki1989, Tasaki1990}. Thus, it is clear that action potentials possess a mechanical component.

To simulate colliding pulses we applied the soliton theory that considers the nerve pulse as an electromechanical compressional pulse. It makes use of the hydrodynamic theory of sound propagation in the presence of nonlinear materials in the presence of dispersion. The nonlinearity in the elastic constants is generated by a phase transition in the lipid chains that influences the elastic properties of the membrane. The soliton theory has the following features: It describes an adiabatic pulse in a membrane cylinder (the axon) in which by necessity no heat is dissipated. Thus, the temperature of a nerve would be the same before and after the pulse \cite{Heimburg2005c}. This has in fact been observed in numerous experiments \cite{Abbott1958, Howarth1968, Ritchie1985}. During the pulse, a change in both nerve area and thickness is predicted. This has been confirmed in early experiments that find both a contraction of the neuron and a slight dislocation of the membrane by about 1 nm (e.g., \cite{Iwasa1980a, Iwasa1980b}). In contrast, the Hodgkin-Huxley model is of a dissipative nature and should result in measurable changes in heat that are not found in experiments. Further, since neither mechanical dislocations or temperature changes are explicitly contained in the Hodgkin formalism, it cannot be used to describe them. It is interesting to note that the soliton theory also contains a feature comparable to a refractory period \cite{Villagran2011}. It is the consequence of mass conservation. The action potential in the soliton theory consists of a region of higher area density of the neuronal membrane. To obey mass conservation, each pulse must be accompanied by a dilated region that prevents that pulses can be arbitrarily close. However, the existence of such a feature does not prevent pulses from penetrating nearly without dissipation.

The earthworm axon consists of many single neurons connected by gap junctions, and one may not consider it as representative for single axons of other species.  We note, however, that the action potential in the earthworm is larger than the dimension of the individual neurons in the axon. Thus, the pulse is a property of the axon as a whole and not of the individual neurons. Further, we provided evidence for the giant axons of the ventral cord of lobster that suggests that observation of undistorted penetration of action potentials is more generic.  

\begin{acknowledgments}
We thank to Prof. Andrew D. Jackson from the Niels Bohr International Academy for useful discussions and for a critical reading of the manuscript. This work was supported by the Villum Foundation (VKR 022130).
\end{acknowledgments}

%

\end{document}